# Conservative relativity principle: Logical ground and analysis of relevant experiments

Alexander Kholmetskii[1], Tolga Yarman[2], and Oleg Missevitch[3]

[1]Department of Physics, Belarus State University, Minsk, Belarus, e-mail: khol123@yahoo.com
[2]Okan University, Istanbul, Turkey & Savronik, Eskisehir, Turkey
[3]Institute for Nuclear Problems, Minsk, Belarus

**Abstract.** We suggest a new relativity principle, which asserts the impossibility to distinguish the state of rest and the state of motion at the constant velocity of a system, if no work is done to the system in question during its motion. We suggest calling this new rule as "conservative relativity principle" (CRP). In the case of an empty space, CRP is reduced to the Einstein special relativity principle. We also show that CRP is compatible with the general relativity principle. One of important implications of CRP is the dependence of the proper time of a charged particle on the electric potential at its location. In the present paper we consider the relevant experimental facts gathered up to now, where the latter effect can be revealed. We show that in atomic physics the introduction of this effect furnishes a better convergence between theory and experiment than that provided by the standard approach. Finally, we reanalyze the Mössbauer experiments in rotating systems and show that the obtained recently puzzling deviation of the relative energy shift between emission and absorption lines from the relativistic prediction can be explained by the CRP.

## 1 Introduction

Nowadays, modern physics accepts two relativity principles: the special relativity principle (SRP) asserting that fundamental physical equations do not change (they are form-invariant) under transformations between inertial reference frames in an empty space, and the general relativity principle (GRP) stating that fundamental physical equations do not change their form (they are covariant) under transformations between any frames of reference. Both relativity principles were introduced by Einstein at the beginning of 20$^{th}$ century: SRP lays at the basis of special relativity, while GRP in combination with the equivalence principle gave rise to general relativity theory.

A fundamental physical consequence of SRP is the impossibility to distinguish between the state of rest of a given system and the state of its motion at a constant velocity in empty space by means of "internal" measuring procedures, which exclude any signal exchange with the outer world.

The GRP is one of the deepest principles of physics and it means that any phenomenon can be described from any reference frame that can be realized in nature (see, e.g. [1, 2]). Historically, the GRP was introduced for gravitation fields, but it can be well applied to any kind of interaction, for example, to joint action of electromagnetic and gravitation fields. In this case the Maxwell equations written in empty space are replaced by their covariant counterparts (see *e.g.* [1]). It is commonly believed that GRP remains in force in quantum theory, too, in spite of the absence of self-consistent theory of quantum gravity up to date[1].

In the present contribution we consider the case of combining the action of various fields (first of all, gravitational and electromagnetic), where we suggest introducing one more relativity principle as a novel postulate. It can be formulated in the following way:

- it is impossible to distinguish the state of rest of any system and the state of its motion with a constant velocity, if this system receives *no work* during its motion.

---
[1] At the same time, one should mention the popular enough Hořava-Lifshitz model (see, *e.g.* [3]), which admits a breakage of GRP at very short distances.

We call this rule "Conservative Relativity Principle", or CRP in short. One can see that in the case of empty space, the CRP is simply reduced to SRT. Hence, classical mechanics, classical and quantum electrodynamics are well fitted into CRP. However, in our opinion, this observation does not exclude a possible heuristic potential of this principle even in these areas of physics (see *e.g.* [4], where some aspects of this problem, although without the explicit formulation of CRP, have been discussed). Besides, there seems no way CRP may enter in conflict with GRP (see below) and thus, both these principles can complement each other only.

In section 2 we find more logical motivations for introducing CRP, considering, in particular, the motion of a massive charged particle in a superimposed gravitational and electric field. We show that without the introduction of CRP, the energy conservation law is violated for the isolated system "particle plus fields" as seen by local observers, tracking the motion of this particle (sub-section 2.1). A closer look at the energy conservation law for a charged particle located in a superimposed gravitational and electric field (sub-section 2.2) allows resolving this paradox via the assumption about the variation of time rate of a charged particle as the function of electric potential at its location. The latter effect can be subjected to the experimental test, and the relevant experiments on this subject are considered in section 3. Finally, section 4 contains a conclusion.

## 2 Logical motivations for the introducing of "conservative relativity principle" and its general implications

We begin this section with the analysis of the motion of a massive charged particle in a superimposed gravitation and electric field, where in the framework of the common approach (i.e. without introducing CRP) we get the violation of the energy conservation law in terms of local (physical) characteristics of particle's motion (sub-section 2.1). Then, in subsection 2.2 we explictly formulate CRP for the case of superimposed gravitational and electric field and derive some of its implications, in particular, the dependence of time rate of a charged particle on an electric potential at its location.

*2.1. Massive charged object in the static gravitation and electric fields: the common approach*

Here we consider the motion of a charged massive point-like object of proper mass $M$ and charge $e$ in a combined gravitation and electric fields, considering for simplicity a case, where both fields are static, and act on the object so to cancel each other.

In order to determine the balance of these forces in terms of characteristics of fields and the object of concern, first we have to define explicitly the corresponding expressions for gravitational and electric forces, assuming for simplicity the one-dimensional case, where the object is moving along $x$ axis at the velocity $\boldsymbol{v}\{v,0,0\}$, measured by a local fixed observer, and both forces act along the $x$-coordinate solely.

As known, the gravitational force in a static/stationary gravitation field is defined by the equation [5]

$$\boldsymbol{F}_g = \gamma M c^2 \left( -\nabla \ln \sqrt{g_{00}} + \frac{\sqrt{g_{00}}}{c} \left( \boldsymbol{v} \times (\nabla \times \boldsymbol{g}) \right) \right), \tag{1}$$

where $\gamma$ is the Lorentz factor of the object to be measured by a local observer, and the components of vector $\boldsymbol{g}$ are equal to $g_i = -g_{0i}/g_{00}$ ($i=1\ldots3$).

In a static gravitation field the metric coefficients $g_{0i}$ are equal to zero, and eq. (1) takes the simpler form

$$\boldsymbol{F}_g = -\gamma M c^2 \nabla \ln \sqrt{g_{00}}. \tag{2}$$

Correspondingly, the metric of space-time in a static field reads as:

$$ds^2 = g_{00} c^2 dt^2 - g_{ij} dx^i dx^j. \tag{3}$$

In the adopted one-dimensional case, the metric coefficients $g_{\mu\nu}$ ($\mu$, $\nu=0\ldots3$) represent the function of $x$-coordinate alone, so that the gravitational force has just a non-vanishing $x$-component, i.e.

$$(F_g)_x = -\gamma Mc^2 \frac{\partial}{\partial x}\ln\sqrt{g_{00}} = -\frac{\gamma Mc^2}{2g_{00}}\frac{dg_{00}}{dx}. \qquad (4)$$

We point out that the forces (1), (2) and (4) are defined as the covariant derivative of three-momentum of moving object with respect to its proper time.

Next, we determine the electric force on the object, which can be found via the motional equation of charged particle in an electromagnetic field in the presence of gravitation [1, 6]:

$$Mc\frac{Du^\mu}{ds} = \frac{e}{c}T^{\mu\nu}u_\nu, \qquad (5)$$

where $Du^\mu/ds$ stands for covariant derivative, $ds$ is the space-time interval, $T^{\mu\nu}$ is the tensor of electromagnetic field, and $u^\mu$ is the four-velocity of particle.

For the case, where the magnetic field is equal to zero, and electric field is static with a single non-vanishing $x$-component $E$, we have just a single non-vanished component of the tensor $T^{\mu\nu}$ [1]

$$T^{10} = E/\sqrt{g_{00}}.$$

Here we emphasize that in the latter equation the electric field $\boldsymbol{E}$ is defined in the absence of gravitation. In the assumed one-dimensional case, $E = -\partial\varphi/\partial x$, where the scalar potential $\varphi$ is also defined in the absence of gravitation. Taking into account the expression for the zeroth component of four-velocity in gravitation field [1]: $u_0 = \gamma/\sqrt{g_{00}}$, we obtain the electric force experienced by the object of concern in the form of

$$(F_e)_x = -\frac{e}{g_{00}}\frac{\partial\varphi}{\partial x}. \qquad (6)$$

Further on we assume that at the moment $t=0$ the gravitational and electric forces do balance each other, i.e. $(F_g)_x + (F_e)_x = 0$, or, via eqs. (4) and (6),

$$-\frac{\gamma Mc^2}{2e}\frac{dg_{00}}{dx} = \frac{\partial\varphi}{\partial x}. \qquad (7)$$

Thus eq. (7) establishes the relationship between the characteristics of object and characteristics of superimposed static gravitational and electric fields, which provide the balance of gravitational and electric forces in the one-dimensional case.

In the limit of a weak gravitation field, we can introduce the gravitational potential $f(x)$ via equation

$$g_{00} = 1 + 2f/c^2, \qquad (8)$$

and directly find the relationship between gravitational and electric potentials, when the gravitational and electric forces mutually cancel each other:

$$-\frac{\gamma M}{e}\frac{df}{dx} = \frac{\partial\varphi}{\partial x}. \qquad (9)$$

This relationship is obtained through the substitution of eq. (8) into eq. (7), and is valid to the accuracy of calculations $c^{-2}$, which is sufficient for further analysis and is adopted hereinafter.

Eq. (9) implies that the gravitation and electric forces lie in opposite directions. We further assume that the electric force has the negative $x$-component, whereas the gravitation force has the positive $x$-component.

Integrating eq. (9), we get one more relationship

$$f = -\frac{e}{\gamma M}\varphi + const, \qquad (10)$$

and in the gauge, where *const*=0, eq. (10) tells us that the gravitation and electric forces mutually balance each other, when the ratio of gravitation and electric potentials is fixed and, in particular, is equal to the charge-to-mass ratio for the resting particle ($\gamma$=1) with the reverse sign.

We emphasize that eqs. (9), (10) are both valid in the entire field region in question, which implies the identical dependence of gravitational and electric potentials on spatial coordinate ***r*** (here we remind that the electric potential $\varphi$ in eqs. (9), (10) is defined in the absence of gravitation). In particular, we can suppose that the electric field ***E*** is constant. In this case the electric potential represents a linear function of *x*-coordinate, and we find from eq. (10) that the gravitation potential also linearly depends on *x*:

$$f = -\frac{e}{\gamma M} Ex. \tag{11}$$

Since the metric coefficient $g_{00}$ must be positive, the latter equation establishes the restriction to the typical size of the field region $x < \gamma Mc^2/2eE$. In fact, the adopted approximation of a weak gravitation field, which implies the accuracy of calculations $c^{-2}$, makes this inequality much stronger:

$$x << \frac{\gamma Mc^2}{2eE}, \tag{12}$$

which simply means that the term $f/c^2$ in eq. (8) is much smaller than unity, when the gravitation potential is defined by eq. (11).

We further assume that the inequality (12) is always fulfilled, though the size of the field region can be large enough.

Further on we emphasize that for the equality of gravitational and electric forces (9), the covariant derivative of momentum of particle with respect to its proper time $\tau$ becomes equal to zero. The same result is valid for the spatial components of four-velocity of particle, i.e.

$$\frac{Du^i}{d\tau} = \frac{du^i}{d\tau} + \Gamma^i_{jk} u^j u^k = 0, \tag{13}$$

where $\Gamma^i_{jk}$ is the Christoffel symbol. For one-dimensional motion along the axis *x*, the latter equation is transformed into

$$\frac{Du^1}{d\tau} + \Gamma^1_{11}(u^1)^2 = 0, \text{ with } \Gamma^1_{11} = \frac{\partial(\ln\sqrt{-g})}{\partial x},$$

where *g* is the determinant of metric tensor $g_{\mu\nu}$. In the approximation of a weak time-independent gravitation field, where the metric coefficient $g_{00}$ is defined by eq. (8), other diagonal metric coefficients are equal to [6]

$$g_{11} = -(1 - 2f/c^2), \; g_{22} = g_{33} = -1. \tag{14}$$

One can see that for the metric coefficients (8), (14), $g = g_{00}g_{11}g_{22}g_{33} = -1$. Hence $\Gamma^1_{11} = 0$, and the covariant derivative is reduced to usual derivative in eq. (13). Thus, we get the equality

$$du_x/d\tau = 0, \text{ and } dv/d\tau = 0.$$

Here $v=dl/d\tau$ stands for the velocity of particle along the axis *x* measured by a local observer, and *dl* is the line element.

Thus, when the balance of gravitational and electric forces is achieved, the velocity ***v*** of object and the Lorentz factor $\gamma$ associated with it, represent the constant values for local observer, at least in the adopted approximation of the weak gravitation field. Therefore, if the motion of the object is not perturbed by any external factors, the balance of gravitational and electric forces, expressed by eq. (9), must be maintained for the entire time, while the object is moving in the field region in question, and we can characterize such an energy state of the system as stationary for local observers.

Further on we consider the following situation: at some time moment *t* the object *M* collides head-on with a very light electrically neutral point-like particle with a rest mass *m*<<*M*, being practically at an infinite distance from the object *M* at the initial time moment, when the bal-

5ance (9) of electric and gravitation fields has already been achieved, and originally resting on the axis *x*. Thus we are well entitled to neglect the gravitation interaction between the masses *M* and *m* in comparison with the available static gravitation field, so that the equality (9) holds true at any instant, excepting the small time interval, when the distance between *M* and *m* becomes very short.

In such short-range interaction and in the process of collision, the object *M* transmits some portion of its energy and momentum to the particle *m*. Naturally, here we assume the conservation of energy-momentum in this short-range interaction; in particular, the total energy $\Delta E_m$ transmitted to the particle *m* is exactly equal to the energy $\Delta E_M$ lost by the particle *M* through the collision. For the purpose of the present paper there is no need to determine exactly these quantities. It is only important to notice that the transmitted energy and momentum have non-vanishing values already in the non-relativistic limit (in particular, in this limit the velocity of particle *m* along the axis *x* becomes twice of the velocity of particle *M* just after the collision), so that the adopted accuracy of calculations $c^{-2}$ is quite a legitimate for further analysis.

Thus, after the collision the particle *m* is rapidly moving away from the object *M*, and during a short time interval their gravitation interaction again becomes negligible, and eq. (9) again is applicable. Designating by $\gamma'$ the Lorentz factor of the object *M* after the collision, we get evidently the inequality $\gamma' < \gamma$. This inequality destroys the balance of electrical and gravitational forces (9) maintained before the collision, no matter how close the two Lorentz factors may be. Thus just after the collision, the magnitude of electric force (which lies in the negative *x*-direction) becomes larger than the magnitude of gravitation force (lying in the positive *x*-direction), and the resultant force on the object *M* is non-vanishing and has a negative projection onto the *x*-axis. This force leads to a further deceleration of object *M* along the *x*-axis, and further decreases its Lorentz factor and, accordingly, the magnitude of the gravitation force (4), acting on it. In its turn, this effect aggravates the misbalance of gravitational and electric forces and enhances the related deceleration of the object *M*. Such a process continues up to the moment, at which the object *M* stops. At this time moment the gravitational force reaches its minimum; the difference of electric and gravitational forces reaches a maximum, and it begins to accelerate the object *M* in the negative *x*-direction.

In this acceleration process, the Lorentz factor of object *M* is continuously increased, which causes the corresponding increase of gravitational force on *M*. As a result, the difference of electric and gravitational forces becomes lower, which leads to the decrease of acceleration of *M*. Thus, further time evolution of this object is obvious: it continues to accelerate, while its Lorentz factor reaches the value, defined by eq. (9). With this Lorentz factor, the electric and gravitational forces again exactly balance each other, just like they did at the initial time moment *t*=0. The velocity of object is equal to its original velocity *v* (though with the opposite sign), and the system reaches the same stationary energy state, as that it had assumed prior to the collision with the particle *m*. We stress that this result inevitably follows from the fact that the balance of gravitational and electric forces is velocity-dependent and thus does not depend on the prehistory of the particle's motion.

Thus we face a situation, where the particle *m* received some portion of energy in the head-on collision with the object *M*, but the latter after some time reaches a stationary state, delineated by exactly the same energy as that it did bear at the initial time moment. It is clear that this result contradicts the energy conservation law with respect to local characteristics of motion[2].

One can add that the electrically charged object *M* emits some portion of electromagnetic radiation in its collision with particle *m* and in further deceleration/acceleration processes. How-ever, the force of reaction of radiation and the Schott term [7, 8] have the order of magnitude $c^{-3}$ and both can be ignored in our analysis, implemented to the accuracy $c^{-2}$.

---

[2] Here it is important to remind that the maintenance of constant electric field acting on a moving charged particle does not require, in general, supplying an external energy to the system. For example, one can imagine that the constant electric field is created by a sufficiently large parallel plate charged capacitor, where the charges are homogeneously distributed over the surface of insulator plates and remain fixed during the motion of the charged particle.

Thus we have shown that for a stationary motion of the object in such a field, when gravitational and electric forces are both collinear to the motional trajectory and mutually balance each other (stage 1), an external perturbation of this motion (in our case due to the collision with some particle) causes the deceleration of this object along its motional trajectory up to the moment, when the velocity of the object is equal to zero (stage 2). Then the object begins to move in the reverse direction with a non-zero acceleration (stage 3), and finally reaches a stationary state with the same value of velocity (as measured by local observers) and the same energy, like in the original stage 1 (stage 4). This effect of "self-stabilization" of the system "moving object in the static electric and gravitational fields", described with the sufficient accuracy of calculations $c^{-2}$, contradicts the energy conservation law, formulated for a set of local observers. We remind that just the local characteristics of motion are recognized as the true (physical) quantities in general relativity, which make the revealed contradiction with the energy conservation law especially strong.

We see just one way to eliminate this contradiction: to re-consider the motion of charged particle in the combined gravitational and electric field with the explicit introduction of CRP, next sub-section.

*2.2. Massive charged object in the static gravitation and electric fields and conservative relativity principle*

Thus, let us again consider a classical point-like change (electron) with the charge $e$ and rest mass $M$ located in the static homogeneous electric and gravitation fields. Hereinafter we again assume the weak relativistic limit, which simultaneously implies a weak gravitation field [5], where the Newtonian limit of general relativity theory becomes relevant. These adoptions allow us to simplify substantially the mathematical side of further analysis and to focus our attention to its principal physical consequences. In the adopted limit, the homogeneous gravitation field is characterized by the gravitation potential of eq. (8), while the electric field is described by the electric potential $\varphi_g = \varphi/\sqrt{g_{00}}$ in the presence of static gravitation field [5], where $\varphi$, as before, being the electric potential in the absence of gravitation field. Further on we again assume the balance of gravitational and electric forces, expressed by eq. (10), and consider the electron at rest in a frame of observation. Thus $\gamma=1$, and in the appropriate gauge equation (10) yields
$$Mf = -e\varphi. \tag{15}$$
for each spatial point of a rest frame of the electron. We stress that eq. (15) is valid only within the adopted approximation, which implies the accuracy of calculations $c^{-2}$, and a general analysis of the combined action of gravitation and electric fields on charged particles in the covariant form will be done elsewhere.

Further, it is well known that the energy of a resting particle in the gravitation field is determined by the relationship [5]
$$E_g = Mc^2 \sqrt{g_{00}}. \tag{16}$$
In the weak relativistic limit the coefficient $g_{00}$ is defined by eq. (8), and $\sqrt{g_{00}} \approx 1 + f/c^2$. Hence
$$E_g \approx Mc^2(1 + f/c^2) = Mc^2 + Mf. \tag{17}$$
within the adopted accuracy of calculations.

Now consider the displacement of the electron from one spatial point *r* to another point *r+dr* within the region defined by eq. (15). Since in this region the gravitational and electric forces mutually cancel each other, *no work* is done to the electron during this replacement. Hence *the energy conservation law* yields:
$$dE_g = -dE_e, \tag{18}$$

where we designated $E_e$ the energy of electron in the electric field. Combining eqs. (18), (17), (15), we derive:

$$dE_e \approx -d(Mf) = d(e\varphi). \quad (19)$$

The obtained equation has a simple physical meaning, if we recall that the term $e\varphi$ represents the interaction electric energy of the electron and a source of the external field, which should be added to the total rest energy of the electron[3]. Thus the total energy of electron in combined gravitation and electric fields is equal to (see eqs. (15), (17), (19))

$$E_{total} \approx Mc^2 + Mf + e\varphi = Mc^2, \quad (20)$$

i.e. it remains constant at least within the adopted accuracy of calculations $c^{-2}$.

Further on we recall a direct relationship existing between the quantities "energy", "frequency' and "time rate" (see, *e.g.* [10]). Thus, considering again the displacement of the electron from the point $r$ to the point $r+dr$, it seems unphysical to admit a change of its proper time rate in these points for the overall fixed total energy (20). Moreover, any assumption on a variation of a proper time rate for an object with a fixed energy directly contradicts the energy conservation law, if one considers, for example, a charged system with the quantized energy levels. Indeed, if we adopt that the displacement of this system within the region (15) induces a variation $\delta\tau$ of the rate of its proper time $\tau$, then the corresponding correction should be introduced to the energy levels $\delta E = h\delta\nu$, where $h$ being the Planck constant, and $\nu$ is the related frequency of radiation (so that $\delta\nu = \nu/(1+\delta\tau/\tau)$). Since the latter relationship is applied to every discrete energy level, it means a consecutive proportional change of the entire energy of the system in question in a superimposed gravitational and electric field. However, this conclusion directly contradicts the constancy of the energy of the system in such a field, determined by eq. (20). Thus, we happen to be in the obligation of assuming that the proper time rate for the electron (or any charged particle that we would operate with, based on eq. (15)) does not depend on the spatial coordinate $r$. On the other hand, it is known that in a static gravitation field, a time rate at any fixed spatial point is determined by the equation

$$d\tau_g = \sqrt{g_{00}}dt \approx (1+f/c^2)dt, \quad (21)$$

where $t$ is the world time (defined, for example, for a distant fixed observer located outside the field region). Thus, in order to keep the total time rate of electron unchanged, we have to adopt that the electric field itself also influences the electron's time rate $\tau_e$, in order to counteract the gravitational effect (21). One can see that if we have

$$d\tau_e = (1+e\varphi/Mc^2)d\tau_g, \quad (22)$$

then the total time rate of electron in the combined gravitation and electric field can indeed remain unchanged:

$$d\tau_{total} = (1+e\varphi/Mc^2)(1+f/c^2)dt \approx dt, \quad (23)$$

at least within the adopted accuracy of calculations $c^{-2}$.

Analogously, one can show that the line elements in the electric field also change:

$$dl_e = dl_g(1+e\varphi_g/Mc^2), \quad (24)$$

where $dl_g$ is the line element in gravitational field the absence of electric field. This effect counteracts the corresponding change of line elements in the gravitation field [6], so that

$$dl_{total} = dl. \quad (25)$$

Now collecting eqs. (20), (23) and (25) altogether, we have:

$$E_{total} = Mc^2, \ d\tau_{total} = dt, \ dl_{total} = dl. \quad (26a\text{-}c)$$

We see that this combination signifies the equivalence of all points $r$ for the electron in question within the spatial region defined by eq. (15). Hence the implementation of CRP for this electron becomes trivial.

---

[3] Herein it is worth to remind a recent paper by Antippa [9] who has shown that for the motion of a test (light) charge in the electric field of a host (heavy) charge, the entire interaction energy must be attributed to the former.

When the electric field exists alone (and this will be implied our further research), eqs. (20), (22) and (24) remain in force with the replacement $d\tau_g \to dt$, $dl_g \to dl$ (i.e. in the absence of gravitational field). Hence, collecting these equations altogether, we obtain:

$$E_e = Mc^2(1 + e\varphi/Mc^2), \quad d\tau_e = (1 + e\varphi/Mc^2)dt, \quad dl_e = (1 + e\varphi/Mc^2)dl. \quad (27\text{a-c})$$

Let us assume a vanishing electric potential $\varphi$ at the infinity. Then in the classical limit we get the equality $e\varphi(P) = \int_{\infty}^{P(r,\vartheta,\varphi)} \mathbf{F} \cdot d\mathbf{s}$, where $\mathbf{F}$ is the force experienced by the particle, $s$ is the displacement, $r$, $\vartheta$, $\varphi$ are the spherical coordinates, so that $\int_{\infty}^{P(r,\vartheta,\varphi)} \mathbf{F} \cdot d\mathbf{s}$ is the work done by this force to bring the particle from the infinity to a given location $P$. Hence eqs. (27a-c) can be rewritten in the form

$$E_e = Mc^2\left(1 + \frac{1}{Mc^2}\int_{\infty}^{P(r,\vartheta,\varphi)} \mathbf{F} \cdot d\mathbf{s}\right), \quad d\tau_e = \left(1 + \frac{1}{Mc^2}\int_{\infty}^{P(r,\vartheta,\varphi)} \mathbf{F} \cdot d\mathbf{s}\right)dt, \quad dl_e = \left(1 + \frac{1}{Mc^2}\int_{\infty}^{P(r,\vartheta,\varphi)} \mathbf{F} \cdot d\mathbf{s}\right)dl$$

(28a-c)

for the location $P$, which simultaneously represents a generalization of eqs. (27a-c) to the case of combined action of electric and gravitation fields.

Eqs. (28a-c) have been suggested originally by Yarman in refs. [11-13], and they are straightforwardly related to the basic formulation of CRP.

In a further analysis we deal with the electric fields only, where eqs. (27a-c) define the change of metrics of space-time in the electric field[4]. Herein one needs to point out an essential qualitative difference between electromagnetic (EM) and gravitation fields. The latter has a universal character, whereas EM interaction involves only electrically charged particles. This means, in particular, that an electrically neutral particle located near the electron does not "sense" the variation of metrics with electric potential (27b-c). Formally this can be taken into account by putting $e=0$ in these equations. An important observation follows from there: eqs. (27b-c) for the temporal and spatial intervals reflect not only the properties of space-time in the electric field, expressed via the scalar potential $\varphi$, but also the properties of particle (the ratio $e/M$), and only the product $e\varphi/Mc^2$ is relevant. This allows assuming the presence of some specific dynamical processes in the interaction of elementary charged particles (for example, electrons, muons) with vacuum, which determine not only a sort of particle, but also are characterized by the sensitivity of such processes to the time rate and spatial scales to be affected by the external electric field.

Classically, the dependence of temporal (27b) and spatial (27c) intervals on the ratio $e/M$ signifies the metric change in a single point of four-dimensional space-time belonging to the world line of the point-like charged particle in question. In the entire free of charges space, the metrics of space-time practically remains Minkowskian (in the absence of gravitation fields), which once again indicates on the compatibility of CRP with SRT and GRP. We can add that within the quantum domain, the metric change of space-time in the electric field can be extended to a vicinity of charge, determining the region of its interaction with vacuum, although in this case we get a problem to introduce an electron's proper reference frame, as implied by classical physics. Now we remain within the classical approach and introduce into consideration a special reference frame $K_e$ co-moving with the point-like charged particle $e$, wherein the time rate in the entire space is determined by eq. (27b). One can see that with respect to a "point-like" observer attached to a charge, such a reference frame could be named "synchronous frame" [5], for the equalities $g_{00}=1$, $g_{0\alpha}=0$ ($\alpha=1\ldots3$) are fulfilled.

---

[4] Here one should recall that in general relativity, an electromagnetic field already affects the metric of space-time, because the electromagnetic energy-momentum tensor is included into the source part of the Einstein equation (see, e.g. [5]). At the same time, such an influence of electromagnetic field on metric of space-time, resulting from the Einstein equation, is well negligible in the laboratory scale experiments, which are considered in section 3. Besides, the metric relationships (27a-c), resulting from CRP, include the charge of particle $e$ and, unlike the solutions of Einstein equation, are not extended to the entire space-time, see the following comments in the text.

We will explore the metric properties of the synchronous frame $K_e$ in the case of empty space, where the gravitation field is absent, and the electric field is constant in the considered region of space. First consider the case, where the frame $K_e$ is at rest with respect to an inertial (laboratory) frame K, where the temporal $dt$ and spatial $dl$ intervals, entering into the *rhs* of eqs. (27b-c), are measured. Assuming the Minkowskian metrics of space-time in the frame K (which, as we mentioned just above, implies the absence of gravitation field), we conclude that the transformations (27b-c) do not affect a type of geometry in $K_e$ (it remains pseudo-Euclidean), but transform the Minkowskian metric tensor into oblique-angled metric tensor [2].

In the case, where the electric field/potential varies with time, the metric of four-space in the synchronous frame $K_e$ also becomes time-dependent. One should emphasize that analyzing eqs. (27a-c), we may consider only slow variations of the electric field (*i.e.*, static or quasi-static cases), since for an arbitrary dependence of the external electric field on time, one has to take into account a radiative EM field, where, in general, the applicability of eqs. (27a-c) needs an additional analysis. However, such an analysis falls outside the scope of the present paper.

When the synchronous frame $K_e$ is moving with the constant velocity $v$ with respect to K within the region of a constant electric field, eqs. (27a-c) are properly modified in the frame K as

$$E_e = \gamma Mc^2\left(1+e\varphi/\gamma Mc^2\right), \quad d\tau_e = \gamma\left(1+e\varphi/Mc^2\right)dt, \quad (dl_e)_\perp = \left(1+e\varphi/Mc^2\right)dl_\perp,$$
$$(dl_e)_{//} = \gamma\left(1+e\varphi/Mc^2\right)dl_{//}, \qquad (29\text{a-d})$$

where $dl_\perp$, $dl_{//}$ are the elements of length, fixed in the frame K, which are respectively orthogonal and parallel to the vector $v$.

One can consider the case, where the frame $K_e$ moves with respect to K with some non-vanished acceleration. However, in such a case we have to involve the radiation loses of the accelerated charge $e$, where, in general, eqs. (27)-(29) are no longer valid. The analysis of the general case (a combination of radiative and non-radiative EM fields) will be considered elsewhere.

Thus, further we adopt either the case of a weak electric field, where the radiation of charges is negligible, or some special cases, where the radiation is forbidden at all, *e.g.*, the quantum systems in stationary energy states, considered, in particular, in sub-section 3.1.

Within the adopted restrictions, eqs. (27a-c) and (29a-d) are valid in the non-radiative EM fields with the non-vanished electric and magnetic field components. At the same time, these equations, being applied to point-like charges, do not involve the vector potential (at least within the weak relativistic limit and time-independent fields) due to the known fact that the magnetic field does not influence the energy of charges. One should only remember that the magnetic field must be weak enough, so that the acceleration of charges must remain weak, too, where the radiation is indeed negligible. The general case of time-dependent fields, as we have mentioned above, will be considered elsewhere.

Having determined the scope of validity of eqs. (27a-c) and (29a-d), as well as the restrictions to their application, it is worth commenting on each of these equations separately. Although eq. (27a) for the energy of particle in the electric field is well-known, in the present context it acquires the additional interpretation related to the metric change of the energy in the synchronous frame $K_e$. Here it is worth to emphasize that CRP does not imply any changes in the structure of electromagnetic energy-momentum tensor, because this relativity principle specifies the properties of charged particles in EM fields (eqs. (27), (29)), rather than any modification of the electric and magnetic fields themselves. Since the matter tensor and EM energy-momentum tensor both enter into the Einstein equation of general relativity, the latter also is not modified by CRP, which once more indicates on the compatibility of general relativity principle and CRP.

At the same time, the involvement of CRP to the problems, dealing with the motion of particles in gravitation field, can essentially influence the obtained solutions. In order to demonstrate the validity of this assertion, we again return to the problem of sub-section 2.1 (the motion of charged particle in a superimposed gravitational and electric field), and point out that with eqs. (22) and (23), the stationary energy state of charged particle is never achieved for a set of local observer in the synchronous frames, tracking the motion of this particle under the condition (9), expressing the balance of electric and gravitational fields. Correspondingly, no contradiction with the energy conservation law, (to be found in sub-section 2.1 without the involvement of CRP) is obtained.

Continuing the analysis of eqs. (27a-c), we point out that the change of time rate (27b) is directly related to the change of total energy (27a), which, in turn, is determined by the variation of interactional EM energy. We further emphasize that the dependence of line spatial element on electric potential (27c) occurs in the synchronous frame $K_e$ only, and for the laboratory observer represents an apparent effect caused by the variation of proper time rate (27b) in $K_e$. Further, involving the fact we established above, that the metric change of energy of charge in the electric field (27a) bears only an interpretation difference from the conventional definition of energy of charge in classical electrodynamics, we draw that only eq. (27b) (or its more general form (29b)) describing the change of time rate of charged particle in an EM field, can be subjected to the experimental test. We emphasize that any experiment, looking for this effect, can be considered simultaneously as the test of CRP. In the next section we analyze some modern experimental data, which, in our opinion, firmly support CRP.

## 3 Conservative relativity principle and its experimental verifications

In this section we subsequently consider various experiments, where the change of time rate of charged particle in an EM field, implying by CRP, represents a key factor in their interpretation.

*3.1. Precise physics of hydrogen-like atoms*

It seems at the first glance that the experiments in this area of physics reject CRP with its prediction (27b) about the influence of electric potential on a time rate of charged particle. Indeed, for the electron bound to the proton (the hydrogen atom), the ratio $e\varphi/mc^2 \approx Z\alpha^2$ (hereinafter $m$ stands for the electron's rest mass, and $\alpha$ is the fine structure constant), so that the related charge of the energy of electron, being added to its EM energy, should manifest itself already at the level of fine structure of hydrogen. At the same time, such specific energy shifts, being additional to the spin-orbit coupling, were never observed. This looks like a strong objection against CRP, which, perhaps, explains why this relativity principle was not advanced to the moment.

However, in a series of our recent papers [14-19] we have found that the actual situation in the atomic physics is not so simple. Namely, we paid attention on the fact that the electrically bound particles in the stationary energy states do not radiate, i.e. their EM field consists of the bound (velocity-dependent) component alone. At the same time, basic equations of atomic physics (both in relativistic quantum mechanics and QED), being constructed by analogy with the appropriate classical equations, ignore, however, the principal difference between EM fields of classical bound charges (which generate both bound and radiative field components) and quantum bound charges (whose fields in the stationary states contain the bound component solely).

The neglect of such a difference can be clearly seen in the Breit equation and Bethe-Salpeter equation [20] for the quantum two-body problem, which essentially use the classical analog of the law of conservation of total momentum in the system "particles and fields", expressed as

$$\bm{p}_m = -\bm{p}_M, \qquad (30)$$

where $\bm{p}_m$, $\bm{p}_M$ are the generalized momenta of particles $m$ and $M$, correspondingly, moving in their common EM field.

However, due to the difference in the structure of fields for bound classical and quantum particles, eq. (30) cannot, in general, be straightforwardly extended from the classical to quantum domain due to the known fact that EM radiation which possesses a momentum is absent in the quantum case. Thus, as minimum, *in the quantum equations the generalized momenta $\bm{p}_m$ and $\bm{p}_M$* should be re-defined in the way, which maintains the total momentum conservation law in the absence of momentum component, associated with an EM radiation.

This conclusion represents one of the root points of our approach, named as Pure Bound Field Theory (PBFT), which we believe is reasonable and strong from the physical viewpoint.

In order to re-define $\boldsymbol{p}_m$ and $\boldsymbol{p}_M$ in an appropriate way, in refs. [14, 15] we considered the one- and two-body problems for bound classical charges with the prohibited radiation and proposed to associate such imaginable systems with the classical limits in the description of electrically bound quantum particles. At the moment we see no other way to eliminate the inconsistency mentioned above with respect to quantum counterpart of eq. (30). In addition, there is one more inconsistenly stemming from common atomic physics: the application of inhomogeneous wave equation for the operator of vector potential (whose solution represents the sum of bound and radiative components) to quantum non-radiating systems [19], and we also see no other way for the elimination of this inconsistency behind PBFT.

For the classical one-body problem with the prohibited radiation, the modification of generalized momentum for orbiting electron (aimed to fulfill the total momentum conservation law in the system "particles plus fields" in the absence of radiative field component) is found straightforwardly, and is defined as [15]

$$\boldsymbol{p}_m = \gamma\left(1 + \frac{e\varphi}{mc^2}\right)m\boldsymbol{v}, \qquad (31)$$

where the electric potential $\varphi$ for the bound electron has the negative value. In the quantum limit, this leads to the replacement of the rest mass of electron $m$ by $b_n m$, where the factor $b_n$ is determined by the equation [15]

$$b_n = 1 - \frac{(Z\alpha)^2}{n^2}, \qquad (32)$$

($Z$ is the atomic number, $n$ is the principal quantum number) and, according to eq. (31), the classical limit of this factor is equal to

$$b = 1 + \left(e\varphi/mc^2\right). \qquad (33)$$

Further, it is important to stress that the replacement $m \to b_n m$ is not a sole modification of quantum mechanical equations for the one-body problem (otherwise, we would derive an additional energy shift of levels $n$ in the order $(Z\alpha)^2/n^2$, which were never observed). One more modification is the replacement of the electric interactional energy $U$ between the host charge and orbiting electron by $\gamma U$, where $\gamma$ is the electron's Lorentz factor. At the classical language the replacement $U \to \gamma U$ can be understood on the basis of Heaviside solution of Maxwell equations for the velocity-dependent (bound) electromagnetic field [7], which shows that for a circular motion of classical charge, its electric field at the location of the host charge is orthogonal to electron's velocity and thus, is increased by $\gamma$ times in comparison with the electric field of a resting electron. In the quantum limit we have the replacement $U \to \gamma_n U$, where the factor [15]

$$\gamma_n = \left(1 - \frac{(Z\alpha)^2}{n^2}\right)^{-1/2}. \qquad (34)$$

Further, one can show [14, 15] that the introduction of PBFT factors (32), (34) into the Dirac-Coulomb equation for the quantum one-body problem via the replacements $m \to b_n m$ and $U \to \gamma_n U$ does not influence its common solution. In other words, the factors $b_n$ and $\gamma_n$ mutually cancel each other, and no modification of the gross and fine structure of hydrigen-like atom takes place [14, 15].

For the classical two-body problem with the prohibited radiation of orbiting charges, the modification of generalized momenta of particles $m$ and $M$ in eq. (30) are more comlicated in comparison with the one-body problem, and we address the readers to our papers [15, 16] for more details. In particular, instead of two PBFT factors (32), (34) for the one-body problem, we have to introduce four correcting factors $b_{nm}$, $b_{nM}$, $\gamma_{nm}$ and $\gamma_{nM}$, which depend on the ratio ($m/M$) and $n$, and provide the implementation of the total momentum conservation law for the isolated non-radiating system "particles plus their bound fields". As shown in ref. [15], the introduction of these factors into Breit equation without external field does not affect the gross and fine structure of atomic energy levels; the PBFT corrections (expressed as some combination of factors $b_{mn}$, $b_{Mn}$, $\gamma_{mn}$, $\gamma_{Mn}$) to $nS$ levels and fine structure corrections appear only in order $(Z\alpha)^6 m/M$ [16]; the corrections to hyperfine spin-spin interval, as well as radiative corrections might have the order $(Z\alpha)^2$ [16].

The correction of the order $(Z\alpha)^6 m/M$ to $nS$ levels occurs insignificant for atoms, where $m<<M$ [16]. The exception is positronium ($m=M$), where the PBFT correction to 1$S$-2$S$ interval completely eliminates the discrepancy between calculated and measured data (up to six standard deviations), available up to date [16].

The PBFT correction to 1$S$ spin-spin interval of the order $(Z\alpha)^2$ is again essential only for positronium, where it also allows eliminating the existing discrepancy (about two standard deviations) between calculated and experimental data [16]. It is important to add that for 1S hfs in muonium, the PBFT correction does not emerge [18], so that the available perfect agreement between calculated and experimental data (see, e.g. [21]) remains in force. This result reflects the general feature of PBFT: in all cases (omitted for brevity in the present contribution), where the common results are already in a quantitative agreement with the measurement data, the PBFT correction factors either cancel each other, or give the corrections, lying beyond the measurement precision.

Finally, the PBFT radiative corrections to the atomic energy levels provide the same estimation (though with different uncertainties) for the proton size in the classic 2$S$-2$P$ Lamb shift in hydrogen, 1$S$ Lamb shift in hydrogen, and 2$S$-2$P$ Lamb shift in muonic hydrogen, with the mean value $r_p$=0.841 fm [19], which exactly coincides with the recent measurement of the proton size via the laser spectroscopy in muonic hydrogen [22].

Thus, the success of PBFT in the elimination of available subtle disagreements between theory and experiment in precise physics of simple atoms is actually undoubted.

One more principal prediction of PBFT is the change of time rate of electrically bound particles as the function of electric potential at their location. It stems from the fact that the replacement $m \to b_n m$ for the quantum one-body problem implies the replacement $E = \gamma_n mc^2 \to \gamma_n b_n mc^2$ in the expression for the energy of bound particle, related to its motion. Due to the mentioned above relativistic relationship between the quantities "energy" and "frequency" (or "time rate"), this modification of energy signifies that the time rate $t'$ for bound particle is defined not only via the time dilation effect, expreseed by factor $\gamma_n$, but also via the factor $b_n$, and the time rate for the bound electron, in comparison with the laboratory time $t$, is defined by the equality

$$dt' = b_n dt / \gamma_n . \qquad (35)$$

Based on the physical meaning of the coefficient $b_n$ for one-body problem, determined via its classical limit (33), we get in this limit eq. (29b), derived above as the requirement of CRP. Here we have to stress that in our earlier paper [17] we classified effect (35) as being quantum in its origins, which is not extended to the classical case. However, at that time we still had not advanced CRP, which completely changes this conclusion and indicates the universal character of variation of time rate of charged particle as the function of an electric potential at its location, which is valid both in the classical (eq. (29b) and quantum (eq. (35)) domains. Thus we see a full compatibility between CRP and PBFT; moreover, they require the validity of each other.

Another point is that in the classical world, a variation of time rate of classical charge with the electric potential at its location (eq. (29b)) can be hardly measured in experiments[5]. In contrast, for the electrically bound quantum particles, the same effect, expressed by eq. (35), can be straightforwardly subjected to the experimental test in the case, where the particles are not stable and experience decay (see next sub-section).

---

[5] The experiment of sub-section 3.3 is not at odd with this statement, because the quantum behavior of resonant nuclei occurs essential for the measurement of effect (29b) even if we deal with a macroscopic system, see below.

*3.2. The effect of change of time rate due to electric field: decay rate of bound muon*

A verification of the change of time rate of electrically bound quantum particles can be carried out with meso-atoms, where the negative muon being captured by the atom possesses a property to directly exhibit its time rate via the decay rate $\tau_b$. Since the strength of electric field $E$ and the electric potential are linearly proportional to the atomic number $Z$, we get a unique possibility to observe the dependence $\tau_b(\varphi)$, measuring the decay rate of bound muons in various atoms.

Recently we already analysed the related experiments [17, 19] for the purpose to test PBFT. Now, for the convenience of the readers, we reproduce in a shortened form the results of [19].

The experiments for measurement of decay rate of muons bound to nuclei in various meso-atoms had been carried out in 1960's of the last century [23, 24] and their results at large $Z$ contradict to each other, as well as to the most reliable theoretical predictions by Huff [25], see Fig. 1.

Chronologically, the experiment by Yovanovitch [23] was implemented before the experiment by Blair et al. [24]; moreover a drastic deviation of experimental data of [23] (black points) at large $Z$ from the careful calculations by Huff [25], stimulated the authors of [24] to carry out new measurements on this subject. The results obtained in [24] are shown in Fig. 1 as the hollow circles. One can see that at large $Z$, these results are in agreement with the idealized curve by Huff (thin continuous line in Fig. 1).

Thus, after the implementation of the experiment [24], it was commonly decided that the data by Yovanovitch [23] (black points in Fig. 1) are most likely erroneous, and the entire problem had been supposed to be closed.

However, it is clear that the experimental data *must be compared* not with the idealized curve by Huff (thin continuous line in Fig. 1), but rather with his real curve (bold continuous line in Fig. 1), which is obtained through the corrections of the idealized curve to the difference of electron spectra for bound and free muons, as well as to a finite size of target [25]. We stress that with respect to the real curve, both the Yovanovitch data [23] and Blair et al. data [24] give deviating results. Thus, a crucial question: whose experimental data are incorrect – by Yovanovitch or Blair et al., remained unanswered.

Now, if we adopt the validity of CRP and PBFT, we have to assume that the real curve calculated by Huff (bold continuous line in Fig. 1) is still incorrect, because it does not take into account the change of time rate for a bound muon due to its binding energy, i.e. the factor $b_n$ in eq. (35). Taking into account that the calculations of Huff already include the relativistic dilation of time for a muon, expressed by factor $\gamma_n$ [25], our specific correction to the calculated value of the time rate of bound muon is presented by factor $b_n$ alone. Hence we get the relationship

$$(\tau_{PBFT})_b = b_n (\tau_{Huff})_b, \tag{36}$$

where $(\tau_{Huff})_b$, as the function of $Z$, is presented in Fig. 1 as the bold continuous line.

Thus, using eq. (36), we multiply the Huff data by factor $b_n$, defined by eq. (32) at $n=1$. As the outcome, we obtain the corrected dependence $(\tau_{PBFT})_b$ on $Z$ to be shown in Fig. 1 as the dot line [19]. A similar curve has been obtained in the earlier paper by the third author [26].

We see that at large $Z$ the PBFT curve is in a good agreement with the data by Yovanovitch, and this coincidence makes highly unbelievable that the assumed effect (36) and the measurements by Yovanovitch are both wrong.

Therefore, for the clarification of the entire issue, new experiments for measurement of lifetime of bound muon are highly required, especially for meso-atoms with a large $Z$.

In the next sub-section we consider one more experiment, where the validity of CRP can be tested via the Mössbauer effect in rotating systems.

*3.3. The effect of change of time rate due to electric field: the Mössbauer experiments in rotating systems*

In this sub-section we deal with the Mössbauer effect and consider the resonant nuclei and the electric fields at their location. By definition, at the stationary state of crystal, the average local electric field of crystal on each nucleus is equal to zero: otherwise, the requirement of equilibrium of a crystal lattice is broken[6]. Rigorously speaking, the requirement of vanished local crystal electric field on nuclei is not exact in terrestrial conditions, where one needs to take into account the gravitation field of Earth, exerted on each nucleus. Hence, there appears a counteracting electric field on each nucleus, which, in general, can affect the nuclear time rate. At the same time, the electric potential entering into eq. (27b) can be taken the same for the resonant nuclei in a Mössbauer source and in absorber located at different attitudes. Therefore, no additional relative shift of resonant lines between the source and absorber, induced by the electric field, emerges, and we can measure the gravitation red/blue shift alone, as it was done in the familiar experiment by Pound and Rebka [27].

Quite another situation is realized for the Mössbauer experiments in rotating systems, where the source of resonant radiation is placed on the rotational axis and the absorber is located on the rim of the rotor (or vise versa). These experiments were originally conceived to verify the relativistic dilation of time, which induces the relative second order Doppler energy shift between emission and absorption lines at the value

$$\left(\frac{\delta E}{E}\right)_{rel} = -0.5 \frac{v^2}{c^2}, \quad (37)$$

where $v$ is the tangential velocity of orbiting absorber. For sub-sound values $v \sim 10^{-6} c$, and typical radius of the rotor system $r \sim 0.1$ m, the centrifugal acceleration many orders of magnitude exceeds the acceleration of free fall in the gravitation field of Earth, which thus becomes negligible. Hence in the rotor experiments the electric field on resonant nuclei of the source, located on the rotational axis, can be taken to be equal to zero. In contrast, as seen from the laboratory, the resonant nuclei in the rotating absorber do experience a centrifugal force, and the requirement of equilibrium of the crystal implies the appearance of a local electric field $E_r$ on resonant nuclei, lying in the radial direction and having the value

$$ZeE_r = -m_N \omega^2 r. \quad (38)$$

Herein $Z$ is the atomic number of the resonant atom, $\omega$ is the rotation frequency, and $m_N$ is the mass of the nucleus. Since the electric potential in a crystal lattice is not a well defined quantity, it is more convenient to apply eq. (28b) for the determination of the variation of nuclei time rate in the resonant absorber, where the integral

$$\int \mathbf{F} \cdot d\mathbf{s} = Ze \int_0^r \mathbf{E} \cdot d\mathbf{r} \quad (39)$$

represents the work done to the resonant nucleus for its displacement from the rotational axis to the edge of the rotor. Combining eqs. (28b), (38) and (39), we obtain the relative change of time rate in the local electric field as follows:

$$\frac{\delta t}{t} = -\frac{\omega^2 r^2}{2c^2} = -\frac{v^2}{2c^2}. \quad (40)$$

At the same time, we cannot assert that eq. (40) determines the change of the time rate for the overall nucleus in the electric field. It appears more precise to advance that the electric potential changes the time rate of protons, but does not affect the time rate of neutrons. Such an effect certainly influences the structure of energy levels of the nucleus, but it would be, in general, in-correct to state that the energy of each level varies proportionally to the obtained value $\delta t/t$. Since the energy levels of nuclei cannot be computed up to date, it is also impossible to determine a variation of these levels due to a change of time rate of protons in the electric field. Thus at the qualitative level we may only assume that the variation of energy levels of nucleus should be sensitive not only to the value $\delta t/t$, but also to the fraction of the protons in the nucleus ($Z/A$), *i.e.*

---

[6] Here we omit the fact that for the *s*-electron, its wave function at the location of nucleus is not vanishing, which induces the isomer shift of resonant lines, not considered in this paper.

$$\left(\frac{\delta E}{E}\right)_{\delta t} = F\left(\frac{Z}{A}\right)\frac{\delta t}{t} = -F\left(\frac{Z}{A}\right)\frac{v^2}{2c^2}, \tag{41}$$

where the function $F$ can be determined experimentally, at least in principle.

In the weak relativistic limit, which is perfectly fulfilled in our case, both effects of a relative shift of resonant lines, (37) and (41), are linearly added to each other. Thus a total relative shift of the energy of nuclear level reads

$$\left(\frac{\delta E}{E}\right)_{total} = \left(\frac{\delta E}{E}\right)_{\delta t} + \left(\frac{\delta E}{E}\right)_{rel} = -k\frac{v^2}{c^2}, \tag{42}$$

where we have introduced the coefficient

$$k = 0.5(1 + F(Z/A)). \tag{43}$$

In a rough approximation we can put $F(Z/A) = Z/A$. Hence for the resonant nucleus of iron ($Z$=26, $A$=57), we obtain $k$=0.728. If the relativistic dilation of time exists alone (and by such a way CRP is rejected), that $k$=0.5.

We should like to recall that Yarman *et al*, in 2007, not having taken into account the difference of the effect of electric field on protons and neutrons of nuclei had predicted $k$=1.0 [28].

A series of Mössbauer experiments on a rotor has been carried out at the early 60s of the past century with the goal to verify a relativistic dilation of time for moving objects [29-34], where all of the authors reported a confirmation of the relativistic expression (37). Among these works, a separate attention should be accorded to the experiment by Kündig [29], since he was the only one who applied a first order Doppler modulation of energy of γ-quanta on a rotor at each fixed rotation frequency $\nu$, implementing a motion of the source along the radius of the rotor. By such a way he recorded the shape and the position of resonant line on the energy scale versus the rotation frequency. In contrast, other authors [30-34] measured only the count-rate of detected γ-quanta as the function of $\nu$. Thus, Kündig's experiment is much more informative and reliable than the others. Hence it was especially intriguing for us to find that the data processing of Kündig's experiment was erroneous [35], and after our correction of the errors displayed by Kündig, his own data furnished the coefficient $k$=0.596±0.006. One sees that the deviation of this result from the relativistic value $k$=0.5 exceeds almost 20 times the measuring error and rather supports eqs. (42), (43) than eq. (37).

We would like to emphasize again that due to applied modulation of energy of emitting resonant radiation, Kündig was successful to measure the *position* of resonant line on the velocity (energy) scale, which is insensitive to vibrations of rotor, despite of line broadening. This methodological feature favorably distinguishes the framework of his experiment from other experiments mentioned above [30-34], where an influence of chaotic vibrations on the width of resonant line in fact was ignored. In particular, Kündig observed an approximately exponential increase of the linewidth up to 1.5 times in a full range of variation of the rotation frequency. It does not mean yet that the same appreciable increase of linewidth took place for the rotors applied in [30-34]. At the same time, it is rather difficult to believe that any line broadening was totally absent, as was tacitly assumed by the authors of the mentioned papers [30-34]. Amongst them the experiment by Champeney *et al.* [34] is distinguished by the numerous experimental data, obtained with different absorbers (5 pieces) and Mössbauer sources $^{57}$Co in two different matrices. Via the reanalysis of this experiment, we achieved in [35], we were able to show that the related results are well fitted into $k$>0.5, too. In particular, our estimation with regards to Champeney et al., unlike what they had originally reported, yields $k$=0.61±0.02.

These findings stimulated the performance of our own Mössbauer experiment on a rotor, which depicted neither the scheme of Kündig experiment [29], nor the schemes of other experiments [30-34], in order to get independent information on *k*. In particular, we did not apply the first order Doppler modulation of the energy of gamma-quanta, in order to avoid a direct repetition of the experiment by Kündig. Thus we followed the scheme [30-34], where the count-rate of detected γ-quanta as the function of rotation frequency is recorded. However, in contrast with the experiments [30-34], we did evaluate the influence of vibrations on the measured value of *k*. For this purpose we applied a method, which involves the joint processing of data collected for two selected resonant absorbers with the specified energy shift of resonant lines [36]. As a result, we have got the estimation *k*=0.66±0.03 [36, 37].

Thus, both the old experiments [29, 34] we reanalyzed [35], and our own Mössbauer experiment on a rotor implemented recently [36, 37] *definitely indicate* that the coefficient *k* in eq. (42) is substantially larger than 0.5. At least at the qualitative level, this result proofs the validity of eqs. (43) with its rough estimation of *k*=0.728, we have provided above.

One should pay attention to the deviation of the value of *k* in our experiment (*k*=0.66±0.03) from the results derived for the Kündig experiment (*k*=0.596±0.006) and Champeney experiment (*k*=0.61±0.02), although a conservative average $k_{av}$=0.62±0.02 agrees with the result of each experiment. Nonetheless, further careful experimental research of the Mössbauer effect in rotating systems is highly required, in order to specify more precisely the value of *k* for $^{57}$Fe resonance.

In any case, the result *k*>0.5 strongly supports the validity of CRP.

## 4 Conclusion

In this paper we advanced the conservative relativity principle (CRP) as the generalization of SRT to the cases, where a moving body experiences the action of forces of different nature, so that the resultant force exerted on the body in question is equal to zero and thus no energy is supplied to it, in order to maintain its motion at constant velocity in the inertial reference frame of observation[7]. In this case the CRP states the impossibility to distinguish the state of rest of a body and the state of its motion with a constant velocity.

The case of especial interest is the inertial motion/rest of charged particle in the combined electric and gravitation fields. Herein one should emphasize that CRP represents a *complementary principle* to GRP and, as we emphasized above, it does not mean any modification of the field equations in classical electrodynamics and the structure of general relativity theory. Rather it allows us to describe what happens to the particle, while it is moving in the fields. In particular, we have found that in the synchronous frame $K_e$ co-moving with the charge *e*, the electric field induces the change of metrics expressed by eqs. (27a-c). It is important to notice that a laboratory observer does not detect such a metric change, and hence in the absence of gravitation fields the metric tensor for this observer remains Minkowskian. This coincides with the known result of relativity theory and again indicates the compatibility of CRP with SRP and GRP.

We have found that among the equations (27a-c), describing the change of energy of charge, spatial and temporal intervals in the electric field, correspondingly, only eq. (27b) can be subjected to the experimental test. In this connection we re-analyzed relevant experiments gathered up to now.

We established that in physics of light hydrogenlike atoms, the requirement of CRP about variation of time rate of charged particles in the electric field is naturally implied in the framework of Pure Bound Field Theory (PBFT), which allows eliminating all of the available subtle deviations between theory and experiment in precise physics of simple atoms, and, in particular, yields a proton size of $r_p$=0.841 fm [19], which perfectly agrees with its latest measurements via laser spectroscopy in muonic hydrogen [22].

---

[7] For the combining action of gravitation and, say, EM fields, it is implied that such an inertial frame is attached to a distant enough observer, so that the gravitation field becomes negligible at his location.

A separate attention is given to the analysis of the decay rate of bound muon in meso-atoms for various atomic numbers $Z$. We have shown that the effect (27b), used for the correction of calculations by Huff [25], gives a good agreement with the data by Yovanovich [23] at large $Z$. This observation makes highly unbelievable that the assumed effect (27b) and the measurements by Yovanovitch are both erroneous, though new experiments for measurement of lifetime of bound muon in meso-atoms are required.

Finally, we analyzed the Mössbauer experiments in rotating systems, where a local electric field on resonant nuclei appears due to the centrifugal force, and the related electric potential does change the energy levels of nuclei by eqs. (42), (43). The latter effect, in fact, could be already found in the careful experiment by Kündig, implemented a half of the century ago [29], but was masked by his computational errors, which we corrected [35]. At the present time an additional energy shift for resonant nuclei due to the electric potential was found by ourselves twice: at a first time in the reanalysis of Kündig experiment [35], and at a second time in the performance of our own experiment on a rotor [36, 37].

Finally, we are convinced that the experimental data reanalyzed in this paper, well indicate the validity of eqs. (29a-d) in different kinds of experiments. Simultaneously we point out that the present paper contains only an elementary introduction into the theory, describing a metric change induced by the electric field in the reference frame $K_e$ co-moving with a charged particle, where we focus our attention on a related physics and the experimental confirmation of our basic predictions. A detailed mathematical apparatus, describing the metric change in the synchronous frame associated with a charged particle, moving in an electric field, has been developed recently, for example, in refs. [38, 39]. A covariant description of the combined action of gravitation and EM field without limitations of their strength and level of their variation with time, remains a subject of high interest (especially if CRP will gather new experimental confirmations), and will be achieved elsewhere.

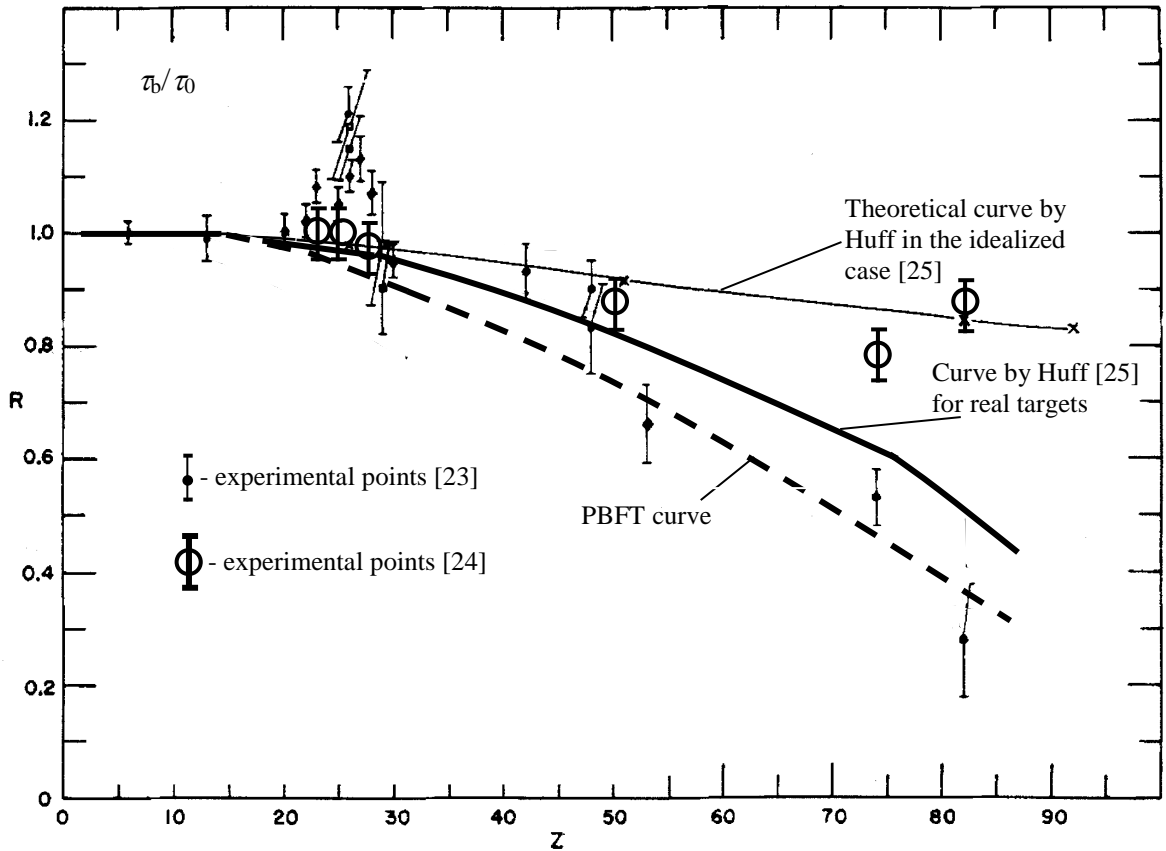

Fig. 1, reproduced from [19]: The dependence of the decay rate $\tau_b$ of bound muon on Z. We compare the results of theoretical calculations by Huff [25] (continuous lines) corrected in PBFT (dot line) with the experimental data of [23] and [24].